\documentstyle[12pt]{article}

\def\qbq{\hbox{$q\bar q$}}
\def\tbt{\hbox{$t\bar t$}}
\def\gmm{\hbox{$g^2m^2_t/M^2_W$}}
\def\mtb{\hbox{$M^2_B$}}
\def\mtt{\hbox{$m^2_t$}}
\def\mth{\hbox{$M^2_H$}}
\def\mtw{\hbox{$M^2_W$}}
\def\mtz{\hbox{$M^2_Z$}}

\def\mft{\hbox{$m^4_t$}}

\def\lsim{\mathrel{\lower4pt\hbox{$=$}}
\hskip-10pt\raise1.6pt\hbox{$<$}\;}

\def\gsim{\mathrel{\lower4pt\hbox{=}}
\hskip-10pt\raise1.6pt\hbox{$>$}\;}

\begin{document}

\setlength{\baselineskip}{18pt}
\begin{flushright}
FERMILAB-PUB-93/027-T\\
BNL-48577\\
\end{flushright}
\vspace*{1truein}
\centerline{\bf YUKAWA CORRECTION TO TOP-QUARK}
\centerline{\bf PRODUCTION AT THE TEVATRON}
\bigskip\bigskip

\centerline{\bf A.~Stange}
\medskip

\centerline{\it Physics Department, Brookhaven National Laboratory,
Upton, NY\ \ 11973}
\bigskip
\centerline{and}
\bigskip

\centerline{\bf S.~Willenbrock}
\medskip

\centerline{\it Physics Department, Brookhaven National Laboratory,
Upton, NY\ \ 11973}
\centerline{\it Fermi National Accelerator Laboratory, P.~O.~Box 500,
Batavia, IL\ \ 60510}
\bigskip\bigskip

\begin{quote}\noindent{\bf Abstract}: We calculate
the correction to $\qbq\to\tbt$ of order $\gmm$. This correction,
proportional to the square of the Higgs-boson Yukawa coupling to the
top quark, arises from loops of Higgs bosons
and the scalar component of virtual vector bosons.
The Yukawa
correction to the total \tbt\ production cross section at the Fermilab
Tevatron in the standard Higgs model is found to be
much less than the theoretical
uncertainty in the cross section.  However, in a two-Higgs-doublet model,
Yukawa couplings are generally enhanced.  The Yukawa correction can
increase the total
$t\bar t$ production cross section in this model
by as much as $20$-$35\%$,
which is potentially observable at the Tevatron.
\end{quote}
\newpage

\leftline{\large \bf 1.\quad Introduction}
\bigskip

In the standard electroweak model, the strength of the weak
interaction is comparable to that of the electromagnetic interaction
at energies of the order of the weak-vector-boson masses. However, the
sector of the weak interaction which generates the fermion masses,
$m_f$, is characterized by a coupling of strength $gm_f/M_W$, where
$g$ is the weak coupling. This is the strength of the Yukawa coupling
of the Higgs boson to the fermions, as well as that of real longitudinal vector
bosons and the scalar component of virtual
vector bosons to the fermions. If the
top-quark mass is much greater than the $W$-boson mass, this
Yukawa-strength coupling is sufficiently large that it could produce
noticeable effects in processes involving the top quark.

The generation of the fermion masses, as well as the weak-vector-boson masses,
is one of the outstanding puzzles of the electroweak theory.
The top-quark coupling to the Higgs boson is a remnant of the mechanism
which generates the top-quark mass.  The observation of effects produced by
this coupling is therefore a window on the top-quark mass-generating
mechanism.  The top
quark could provide us with the first clue towards solving the mystery of mass
generation.

In the standard Higgs model, a single Higgs doublet provides masses to
the fermions and the weak vector bosons.  Consequently, Yukawa couplings
are fixed in terms of the corresponding fermion mass and the $W$-boson mass.
However, in a two-Higgs-doublet model, the Higgs bosons which come
from the doublet which provides mass to a given fermion generally have
enhanced Yukawa couplings to that fermion.  This results in a more
pronounced effect of the fermion mass-generating mechanism on
physical processes.

In this paper we investigate the Yukawa-strength correction to
top-quark production at the Fermilab Tevatron. With the Main Injector,
it may be possible to dectect a top quark as heavy as
250 GeV at the Tevatron. The dominant production
mechanism for a heavy top quark ($m_t>150$ GeV) at the Tevatron
is quark-antiquark
annihilation \cite{BGT}; we therefore restrict our attention to this process.
The diagrams which contribute to this correction are shown in Fig.~1.
The dashed lines
represent the Higgs boson and the unphysical scalar bosons,
in Landau gauge, associated with the vector bosons.
The Yukawa correction to the
top-quark cross section is proportional to \gmm.

In Landau gauge, the vector-boson propagator is entirely spin one, and
the scalar component of the virtual vector boson is represented by the
massless unphysical scalar boson (Goldstone boson).
In this gauge, the interactions of
Yukawa strength are isolated
in the Higgs boson and unphysical scalar boson couplings to the fermions.
Real (but not virtual) longitudinal vector bosons also
effectively couple to fermions with Yukawa strength.

The Yukawa correction to $t\bar t$ production in the standard Higgs model
is discussed in section 2.  In section 3 we discuss the correction in
a two-Higgs-doublet model, both in general and in the minimal supersymmetric
model.  We present our conclusions in section 4.
Analytic expressions for the loop corrections
are given in an appendix.
\bigskip

\leftline{\large \bf 2.\quad Standard Higgs model}
\medskip

The Yukawa correction to the $\qbq\to \tbt$ amplitude (see Fig.~1)
is contained in the
correction to the matrix element of the top-quark color current.  The
general form for this matrix element, consistent with current conservation,
is
\begin{eqnarray*}
i\bar u(p_3)\Gamma^{\mu A} v(p_4)&=&-ig_s \Biggl[\bar u(p_3) T^A \gamma^\mu
v(p_4) \Biggr. \\
&-& \left(\frac{gm_t}{2M_W}\right)^2 \frac{1}{(4\pi)^2} \\
& & \quad\times\;\bar u(p_3) T^A\Big[ V\gamma^\mu +
Ti\sigma^{\mu\nu} q_\nu
+ A(\gamma^\mu q^2-2m_tq^\mu)\gamma_5
+P\gamma_5 \sigma^{\mu\nu}q_\nu
\Big] v(p_4) \Biggr] \\
\end{eqnarray*}
where $p_3$ and $p_4$ are the outgoing momenta of the top and antitop
quark, $q=p_3+p_4$, the form factors $V$, $T$, $A$ and $P$ are functions of
$q^2=s$, $T^A$ is an $SU(3)$ matrix,
and we have factored out the couplings and loop factors.
The chromo-electric-dipole form factor ($P$) is CP violating,
and vanishes at one loop.
The chromo-charge ($V$), chromo-magnetic ($T$), and chromo-anapole ($A$)
form factors are given in an appendix.

The differential cross section for $\qbq\to \tbt$, summed over final and
averaged over initial colors and spins, is
\begin{eqnarray*}
\frac{d\sigma}{dz} &=& \frac{8\pi}{9s^3} \alpha^2_s\beta \biggl[
(p_1\cdot p_3)^2 + (p_2\cdot p_3)^2 + m^2_t p_1\cdot p_2 \biggr. \\
&&\quad -\frac{g^2m^2_t}{64\pi^2M^2_W} {\rm Re} \Bigl[ 2V [ (p_1\cdot p_3)^2
+ (p_2\cdot p_3)^2 + m^2_t p_1\cdot p_2 ] + m_ts^2 T \Bigr] \biggr]\\
\end{eqnarray*}
where $p_1$ and $p_2$ are the incoming momenta of the quark and
antiquark, $z$ is the cosine of the scattering angle between the quark
and the top quark, and $\beta=(1-4m^2_t/s)^{1/2}$. The parity-violating
chromo-electric-dipole
($P$) and chromo-anapole ($A$)
form factors do not contribute to the unpolarized cross section at one loop.

In Fig.~2 we show the percentage change in the cross section, as a
function of the \tbt\ invariant mass, for $m_t=150$ GeV
and $M_H=60$, 200, and 800 GeV\null. The contributions of the Higgs boson and
the unphysical scalar $Z$ and $W$ bosons are shown separately.
The Higgs-boson contribution
near threshold is large and positive for a relatively light Higgs boson.
This is due to the Yukawa
potential formed between the top and antitop quarks
by the exchange of the Higgs
boson, a phenomenon that has been studied in $e^+e^-\to\tbt$ near
threshold \cite{F,SP,GK}. The exchange of the unphysical scalar $Z$ and $W$
bosons does not lead to a similar effect, in the first case because
the interaction is pseudoscalar, in the second because the interaction
changes the top quark to a bottom quark. Exchange of the unphysical scalar
$Z$ boson produces a potential which vanishes in the non-relativistic limit,
so its contribution vanishes at threshold.
The unphysical-scalar-$W$-boson contribution
is negative and almost constant over the range shown. The
Higgs-boson contribution is also negative far above threshold.

To obtain the Yukawa correction to the
top-quark production cross section at the Tevatron, one convolutes the
subprocess cross section with parton distribution functions.
For completeness, one must also consider corrections of order $\gmm$ to the
parton distribution functions. The parton distribution functions
are extracted from deep-inelastic
scattering, Drell-Yan, and direct photon production \cite{OT}. Terms
of order \gmm\ can potentially arise from the top-quark contribution to
weak-vector-boson
vacuum polarization. However, when charged-current processes are
expressed in terms of $G_\mu$, these terms are absorbed into the
coupling, since the same vacuum-polarization diagram occurs in muon
decay. Neutral-current processes do have corrections of
${\cal O}(\gmm)$ via the $\rho$ parameter \cite{V},
\[ \rho = 1 + 3\frac{g^2m_t^2}{64\pi^2M_W^2}\;\;, \]
but presently there are no
neutral-current processes from which information on
the parton distribution functions
are extracted.  Thus there is no Yukawa
correction to the parton distribution functions.

To obtain the correction to the total top-quark production cross
section at the Tevatron, we weight the correction to the subprocess
cross section with the parton distribution functions
and integrate over all \tbt\ invariant masses.
The quark and antiquark distribution functions are evaluated at
$x\sim 2m_t/\sqrt S\sim 0.1-0.2$, where they are well measured.
We present our numerical results with the
Morfin-Tung leading-order parton distribution functions \cite{MT}.
The resulting change in
the total cross section is shown in Fig.~3 for $m_t=150$, 200, and 250 GeV,
as a function of the Higgs-boson mass.
Because the parton distribution functions and the subprocess cross
section fall off with increasing $t\bar t$ invariant mass, only the region
within a few hundred GeV of threshold is numerically significant when we
integrate over $M_{t\bar t}$.
The positive contribution from the Higgs boson is largely compensated by
the negative contribution from the unphysical scalar $W$ boson, so
the correction to the total cross section is much smaller than the
typical correction at fixed invariant mass.  The unphysical $Z$ boson makes
a negligible contribution.  The contribution from a very light Higgs boson
($M_H <$ 100 GeV) is reduced because the Yukawa enhancement is peaked
very close to threshold, where the cross section is suppressed by
phase space. The correction is largest
for $M_H \approx 125$ GeV, but is at most $+2.4\%$ for $m_t<250$ GeV.

The radiation of a real Higgs-boson \cite{NZ,K,BSP} or longitudinal vector
boson \cite{K,BSP} from a top quark is also a correction to top-quark
production of Yukawa strength. We have calculated these processes and
found that they are negligible at the Tevatron. The parton
distribution functions are steeply falling at large masses, so the
emission of an additional massive particle is suppressed.

The uncertainty in the top-quark total cross section at the Tevatron
is due to the uncertainty in $\alpha_s$ and the
uncalculated next-to-next-to-leading order QCD correction \cite{BGT}. The net
uncertainty is estimated to be about $\pm20\%$ \cite{E}, much greater
than the Yukawa correction to the total cross section. We conclude
that the Yukawa correction to the total cross section is unobservable
at the Tevatron for $m_t <$ 250 GeV.

Because the ordinary weak corrections to deep-inelastic scattering,
Drell-Yan, and direct-photon production are not included in the
extraction of the parton distribution functions,
the ordinary weak correction to the $\qbq\to\tbt$ cross
section does not represent a complete calculation of the ordinary
weak correction to the hadronic $t\bar t$ cross section.
This correction, as well as the Yukawa correction which we
have evaluated, was calculated in Ref.~\cite{A}, and numerical results
were given for the correction to the $\qbq\to\tbt$ and $gg\to\tbt$
cross sections. These results agree with ours in the region a few hundred
GeV above threshold, although the threshold region is not shown in
sufficient detail to allow a comparison.
That study notes that there
is a large negative correction to the cross sections
at very large \tbt\ invariant masses. Since we do not find such a result
from the Yukawa correction, it is presumably due to the ordinary weak
correction, enhanced by a large infrared logarithm, $\ln^2 M_V^2/s$.
However, the large-invariant-mass region makes a negligible contribution
to the total hadronic cross section.  We question the estimate of a
negative $10$--20\% correction to the \tbt\ production cross section
at the LHC for $m_t=$ 200 GeV
quoted in that work. At the LHC/SSC $gg\to\tbt$ is the dominant
top-quark production mechanism, so we cannot give a result for the
Yukawa correction at these machines.

The $W$-gluon fusion process, $Wg\to t\bar b$, is also a copious source
of top quarks at the Tevatron, especially for large $m_t$ \cite{DW}.  However,
once cuts are made in an effort to identify the signal, the rate for
top-quark production via $W$-gluon fusion falls below that of
$q\bar q \to t\bar t$ \cite{Y,EP}.

The full weak correction to $e^+e^-\to t \bar t$, including the Yukawa
correction, is given in Ref.~\cite{BVH}.  The correction is comparable
to the correction found here for fixed $t\bar t$ invariant mass.  The
correction in the threshold region is studied in Ref.~\cite{BH}.

\bigskip

\leftline{\large \bf 3.\quad Two-Higgs-doublet model}
\medskip

The simplest extension of the Higgs sector of the standard electroweak model
that does not conflict with experiment is a two-Higgs-doublet model \cite{HHG}.
The tree-level relation $\rho=M_W^2/(M_Z^2\cos^2\theta_W)=1$ is satisfied
automatically, as in the one-doublet model \cite{RV}.  In order to avoid
Higgs-boson-mediated tree-level flavor-changing neutral currents, it is
necessary to require that fermions of a given electric charge receive their
mass from only one of the Higgs doublets \cite{GW}.  Since $M_W^2=\frac{1}{4}
g^2(v_1^2+v_2^2)$, both $v_1$ and $v_2$ must be less than $v$, the
vacuum-expectation value of the one-doublet model.  The Yukawa couplings
of a given fermion to the Higgs scalars which come from the doublet $\phi_i$
that provides mass to that fermion are
proportional to $m_f/v_i$, and are therefore naturally enhanced.
Thus one may obtain an enhanced Yukawa correction to top-quark production
in a two-Higgs-doublet model.

The two-Higgs-doublet model has five physical Higgs bosons, in addition to
the unphysical scalar $Z$ and $W$ bosons. There are two scalars, $h$ and $H$;
a pseudoscalar, $A$; and two charged scalars, $H^{\pm}$.  With regard
to their coupling to the top quark, the pseudoscalar and charged scalars
are massive versions of the unphysical scalar $Z$ and $W$ bosons, respectively,
but with the coupling factors given in Table 1. It is conventional to
chose $\phi_2$ to give mass to the top quark, and to define
$\tan \beta=v_2/v_1$;
the top-quark Yukawa couplings are therefore
enhanced for small $\beta$.  The two scalar Higgs bosons
have the same quantum numbers, and mix with a mixing angle $\alpha$.
The couplings of the unphysical scalar $Z$ and $W$
bosons are unchanged from the one-doublet model.

If $\beta$ is sufficiently small, processes mediated by charged Higgs
bosons may be enhanced such that they conflict with experiment.  The
strongest constraint appears to come from the lack of observation of
$b\to s\gamma$ \cite{CLEO,H}.
For small $\beta$, the upper limit on $b\to s\gamma$
requires a charged-Higgs-boson mass greater than several hundred GeV,
depending on the model and the top-quark mass \cite{H}.
To avoid a large correction to the $\rho$ parameter
we must set $M_A \approx M_{H^\pm}$ \cite{T}, so the pseudoscalar Higgs
boson must also be heavy.

A sufficiently small value of
$\beta$ also yields a Yukawa coupling which is so strong that perturbation
theory is unreliable.
Consider the zeroth partial wave of the $t\bar t$ elastic scattering
amplitude, keeping only the terms which are enhanced for small $\beta$
\cite{CFH},
\begin{eqnarray*}
a_0=-\frac{3G_Fm_t^2s}{8\pi\sqrt 2\sin^2\beta}
\Biggl[\frac{\cos^2\alpha}{s-M_h^2}+\frac{\sin^2\alpha}{s-M_H^2}
+\frac{\cos^2\beta}{s-M_A^2}\Biggr].\\
\end{eqnarray*}
Applying the unitarity condition
${|\rm Re}\;a_0| < 1/2$ in the energy regime $M_{h,H}^2 << s << M_A^2$
implies \cite{MVW}
\begin{eqnarray*}
\sin^2\beta > \frac{3G_Fm_t^2}{4\pi\sqrt 2}.
\end{eqnarray*}
This gives $\beta >$ 0.21, 0.28, 0.36 for $m_t=$ 150, 200, 250 GeV,
respectively. Although $\beta$ may be less than these values in principle,
we believe that they correspond to the largest Yukawa correction that
is physically allowed.

In order to ascertain the maximum Yukawa correction to top-quark production
at the Tevatron that one can expect in a two-Higgs-doublet model, we set
$\beta$
to its minimum value and set $M_{H^{\pm}}=M_A=$ 600 GeV.
The contributions of the
Higgs bosons to the form factors are given by the same expressions as in
the one-doublet model, but with $M_H \to M_{h,H}$, $M_Z \to M_A$, and
$M_W \to M_{H^{\pm}}$, and multiplied by the square of the associated
coupling factor in Table 1.
Since the couplings of the unphysical scalar bosons
are not enhanced, their contribution is negligible.  As with the unphysical
scalar $Z$ boson in the one-doublet model, the pseudoscalar-Higgs-boson
contribution is negligible, for all $M_A$.
Recall that in the one-doublet model the positive contribution of the Higgs
boson was largely compensated by the negative contribution of the unphysical
scalar $W$ boson.
For small $\beta$, the charged Higgs boson plays the role of the
unphysical scalar $W$ boson;
however, since it must be heavy, its
(negative) contribution is suppressed.  Thus, as long as both Higgs scalars
are relatively light, or there is one light scalar whose coupling to the top
quark is not suppressed by the mixing angle $\alpha$, we can expect a large,
positive Yukawa correction in the two-Higgs-doublet model for small $\beta$.

We show in Fig.~4 the maximum
correction to the $t\bar t$ total cross section at
the Tevatron, for $M_h=M_H$, as a function of the common scalar-Higgs-boson
mass. The mixing angle $\alpha$ drops out for equal scalar masses.  The
correction
is increased considerably over the standard-model correction, as large
as $+23\%, +28\%, +35\%$ for $m_t=150$, 200, 250 GeV. These corrections
to the total cross section
are potentially observable at the Tevatron, and could be used to set
limits on the parameters of a two-Higgs-doublet model.

The Higgs sector of the minimal supersymmetric standard model is a special
case of a two-Higgs-doublet model, and is specified by just two parameters
(at tree level), which we can take to be $M_{H^{\pm}}$ and $\beta$ \cite{HHG}.
For large $M_{H^{\pm}}$, $\alpha \to \beta -\frac{\pi}{2}$, so that for small
$\beta$ the $H$ Yukawa coupling is enhanced while that of $h$ approaches
the one-doublet-model value.  For large $M_{H^{\pm}}$,
$H$ is also heavy (it is nearly degenerate in mass with the charged Higgs
boson), so its contribution is suppressed.  Thus we do not find as large
a Yukawa correction to top-quark production at the Tevatron in this model
as in the generic two-Higgs-doublet model.

We show in Fig.~5 the maximum correction to the total top-quark production
cross section at the Tevatron, as a function of the charged-Higgs-boson
mass, in the minimal supersymmetric two-Higgs-doublet model, obtained by
setting $\beta$ to its minimum value for each top-quark mass.
For a light charged Higgs boson, there is a cancellation between the
contribution of the charged Higgs boson and the scalar Higgs bosons,
as mentioned above. The dip
at $M_{H^\pm}\approx m_t$ is due to the rapid variation of the
charged-Higgs-boson form factor at the threshold for $t\to bH^+$.  The implied
lower limit on the charged-Higgs-boson mass from searches for
$Z\to Z^*h$ at LEP is indicated in the figure \cite{LEP}.  The upper
bound on $b\to s \gamma$ \cite{CLEO}
implies $M_{H^\pm}>$ 500 GeV \cite{H}, although
the contribution of supersymmetric particles to this process is not
included in this bound.  The correction to the total cross
section does not exceed $20\%$ for any value of $M_{H^\pm}$, and is
therefore unlikely to be observable at the Tevatron, unless there is
significant progress in the calculation of the total cross section.

The full weak correction to $e^+e^-\to t \bar t$ in a two-Higgs-doublet model,
including the Yukawa
correction, is given in Ref.~\cite{BDK}, and the full weak correction near
threshold is given in Ref.~\cite{DGK}.

\newpage

\leftline{\large \bf 3.\quad Conclusions}
\medskip

We have calculated the correction of Yukawa strength, $\gmm$, to the
total top-quark production cross section at the Fermilab Tevatron.
In the standard Higgs model, the correction is less than $+2.4\%$
for $m_t<$ 250 GeV, much less than the theoretical uncertainty in the
cross section.

The Yukawa correction to the total top-quark production cross section can be
significantly enhanced in a two-Higgs-doublet model, if the doublet
which generates the top-quark mass has a small vacuum-expectation value.
Corrections as large as $+23\%, +28\%, +35\%$ for $m_t=$ 150, 200, 250
GeV are obtained.  Assuming a theoretical uncertainty in the cross
section of $\pm 20\%$ from QCD, these corrections are potentially
observable, and could be used as evidence for, or to place restrictions
on, a two-Higgs-doublet model.  A measurerment of the total cross section
with systematic and statistical errors less than $\pm 20\%$ would be
required.
If one specializes to the minimal supersymmetric two-Higgs-doublet model,
one finds that the correction never exceeds $20\%$.

The top quark may provide a window on mass generation in the weak
interaction.  It is worthwhile to explore the different signatures of the
top-quark mass-generating mechanism in top-quark production at the
Tevatron.  The calculation of the Yukawa correction to the total
top-quark production cross section is a first attempt in this direction.

\bigskip

\leftline{\large\bf Acknowledgements}
\medskip

We are grateful for conversations with W.~Marciano and L.~Trueman, and
for assistance from T.~Stelzer. This manuscript has been authored under
contract no. DE-AC02-76CH00016 with the U.~S.~Department of Energy.
S.~W. was partially supported by an award from the Texas National
Research Laboratory Commission.

\newpage

\leftline{\bf APPENDIX}
\medskip

Below we give the form factors for the matrix element of the top-quark
current. We separately list the contribution from the Higgs boson and
the unphysical scalar $Z$ and $W$ bosons, of mass $M_Z$ and $M_W$
('t~Hooft-Feynman gauge).
We set $M_Z$ and $M_W$ to zero (Landau gauge)
in the numerical calculations, as described
in the introduction.  In a two-Higgs-doublet model, the
pseudoscalar-Higgs-boson
($A$) and the charged-Higgs-boson ($H^{\pm}$) form factors are given by
the unphysical scalar $Z$- and $W$-boson form factors, with $M_Z\to M_A$ and
$M_W\to M_{H^{\pm}}$, and multiplied by the square of the associated coupling
factor in Table 1.  The same is true of the two scalar Higgs bosons
($h,H$) of the two-Higgs-doublet model with respect to the
Higgs-boson form factor.

The form factors are written in terms of the usual one-, two-, and
three-point scalar loop integrals
$A_0$, $B_0$, and $C_0$ \cite{PV}. The integrals were evaluated with the code
FF \cite{FF}, whose
notation we have adopted.

\bigskip

\leftline{\large\bf Higgs boson (scalar Higgs bosons):}

\begin{eqnarray*}
V(s-4\mtt)^2 &=& (s-4\mtt)^2\Biggl[ \frac{1}{2\mtt} [\mth B_0
(\mtt,\mth, \mtt) - A_0(\mth) +A_0 (\mtt)] \Biggr.\\
& & \Biggl. \qquad\qquad + (4\mtt-\mth) B^\prime_0(\mtt, \mth, \mtt)
\Biggr] \\
& & \qquad + [A_0(\mth) - A_0 (\mtt)] [-2(s-4\mtt)] \\
& & \qquad + B_0 (\mtt, \mth, \mtt) [ s\mth - 8s\mtt -16\mtt\mth +32\mft] \\
& & \qquad + B_0 (\mtt, \mtt, s) \left[ -\frac{1}{2} s^2 +s\mth +10s\mtt +8
\mtt\mth - 32\mft\right] \\
& & \qquad + C_0(\mtt,\mtt,\mth,s,\mtt,\mtt) [4s^2\mtt + sM^4_H -32s\mft \\
& & \qquad\qquad + 12s\mtt\mth + 8\mtt M^4_H -48\mft\mth + 64 m^6_t] \\
& & \qquad + \frac{1}{2} s(s-4\mtt) \\
\end{eqnarray*}
\begin{eqnarray*}
T(s-4\mtt)^2 \frac{1}{2m_t} &=& [A_0(\mth) - A_0 (\mtt)]
\frac{1}{2\mtt} (s-4\mtt) \\
& & \qquad + B_0 (\mtt, \mth, \mtt)  \left[-s \frac{\mth}{2\mtt} +
2s + 5\mth -8\mtt\right] \\
& & \qquad + B_0 (\mtt, \mtt, s) \left[-\frac{3}{2} s-3\mth
+6\mtt\right] \\
& & \qquad + C_0(\mtt,\mtt,\mth,s,\mtt,\mtt) [-3s\mth -3M^4_H + 12
\mtt\mth] \\
& & \qquad -\frac{1}{2} (s-4\mtt) \\
\end{eqnarray*}

$\displaystyle A = 0$

\bigskip

\leftline{\large\bf Unphysical scalar $Z$ boson (pseudoscalar Higgs boson):}

\begin{eqnarray*}
V(s-4\mtt)^2 &=& (s-4\mtt)^2 \Biggl[\frac{1}{2\mtt} [\mtz B_0 (\mtt, \mtz,
\mtt) -A_0(\mtz) + A_0(\mtt)] \Biggr. \\
& & \qquad\qquad \Biggl. -\mtz B^\prime_0 (\mtt,\mtz,\mtt)\Biggr] \\
& & \qquad + [A_0(\mtz) - A_0(\mtt)] [-2(s-4\mtt)] \\
& & \qquad + B_0 (\mtt, \mtz, \mtt) [s\mtz-16\mtt\mtz] \\
& & \qquad + B_0 (\mtt, \mtt, s) \left[-\frac{1}{2}s^2 +s\mtz +2s\mtt
+8\mtt\mtz\right] \\
& & \qquad + C_0(\mtt,\mtt,\mtz,s,\mtt,\mtt) [sM^4_Z +4s\mtt\mtz
+8\mtt M^4_Z -16\mft\mtz] \\
& & \qquad + \frac{1}{2}s (s-4\mtt) \\
\end{eqnarray*}
\begin{eqnarray*}
T(s-4\mtt)^2 \frac{1}{2m_t} &=& [A_0(\mtz) - A_0(\mtt)] \frac{1}{2\mtt}
(s-4\mtt) \\
& & \quad + B_0 (\mtt, \mtz, \mtt) \left[ -s\frac{\mtz}{2\mtt} +
5\mtz\right] \\
& & \quad + B_0 (\mtt, \mtt, s) \left[\frac{1}{2}s -3\mtz -2\mtt
\right] \\
& & \quad + C_0(\mtt,\mtt,\mtz,s,\mtt,\mtt) [-s\mtz-3M^4_Z
+4\mtt\mtz] \\
& & \quad - \frac{1}{2}  (s-4\mtt) \\
\end{eqnarray*}

$\displaystyle A = 0$

\bigskip

\leftline{\large\bf Unphysical scalar $W$ boson (charged Higgs boson):}

\begin{eqnarray*}
V(s-4\mtt)^2 &=& (s-4\mtt)^2 \Biggl[ \frac{1}{2\mtt} [(\mtt + \mtw) B_0
(0,\mtw, \mtt) - A_0(\mtw)] \Biggr. \\
& & \qquad  \Biggl. \qquad\qquad+ (\mtt-\mtw)
B^\prime_0(0,\mtw,\mtt) \Biggr] \\
& & \qquad + A_0 (\mtw) [-2(s-4\mtt)] \\
& & \qquad +B_0 (0,\mtw,\mtt) [s\mtw - 3s\mtt-16\mtt\mtw] \\
& & \qquad +B_0 (0,0,s) \left[ -\frac{1}{2} s^2 +s\mtw +7s\mtt
+8\mtt\mtw -8\mft\right] \\
& & \qquad +C_0 (0,0,\mtw,s,\mtt,\mtt) \Bigl[s^2\mtt+sM^4_W - 3s\mft
\Bigr. \\
& & \Biggl. \qquad\qquad +10s\mtt\mtw +8\mtt M^4_W - 16\mft\mtw +8 m^6_t
\Bigr] \\
& & \qquad + \frac{1}{2} s(s-4\mtt) \\
\end{eqnarray*}
\begin{eqnarray*}
T(s-4\mtt)^2 \frac{1}{2m_t} &=& A_0(\mtw) \frac{1}{2\mtt} (s-4\mtt) \\
& & \qquad + B_0 (0,\mtw,\mtt) \left[ - \frac{s\mtw}{2\mtt} +
\frac{1}{2} s + 5 \mtw + \mtt\right] \\
& & \qquad + B_0 (0,0,s) \left[ -\frac{1}{2} s-3\mtw-\mtt \right] \\
& & \qquad + C_0 (0,0,\mtw,s,\mtt,\mtt) \Bigl[ -2s\mtw -s\mtt -3 M^4_W
 + 2\mtt\mtw+ \mft \Bigr] \\
& & \qquad -\frac{1}{2} (s-4\mtt) \\
\end{eqnarray*}
\begin{eqnarray*}
As(s-4\mtt) &=& A_0(\mtw) \frac{1}{2\mtt} (s-4\mtt) \\
& & \qquad + B_0 (0,\mtw,\mtt) \left[ - \frac{s\mtw}{2\mtt} +
\frac{1}{2} s + \mtw + \mtt\right] \\
& & \qquad + B_0 (0,0,s) \left[ -\frac{1}{2} s+\mtw-\mtt \right] \\
& & \qquad + C_0 (0,0,\mtw,s,\mtt,\mtt) \Bigl[ -s\mtt + M^4_W
- 2\mtt\mtw+ \mft \Bigr] \\
& & \qquad +\frac{1}{2} (s-4\mtt) \\
\end{eqnarray*}

\noindent The first two lines of each expression for $V$ is the
contribution from top-quark wavefunction renormalization, with
\begin{eqnarray*}
B^\prime_0 (\mtt,\mtb,\mtt) &=& \frac{1}{\mtt} \left[
\frac{1}{x^B_+ - x^B_-} \left (\frac{(x^B_+)^2}{x^B_-} \ln x^B_- -
\frac{(x^B_-)^2}{x^B_+} \ln x^B_+\right) -1 \right] \\
B^\prime_0 (0,\mtw,\mtt) &=& \frac{1}{\mtt} \left[ r_W \ln
\frac{r_W}{r_W-1} -1 \right] \\
\end{eqnarray*}
where $r_B=\mtb/\mtt\; (B=H,Z,W)$, and
\[ x^B_\pm =\frac{1}{2} [r_B \pm (r^2_B -4r_B)^{1/2}]. \]

\newpage

\leftline{\large\bf Tables}
\medskip
\begin{description}
\item[Table 1.] Factors associated with the top-quark Yukawa couplings
in a two-Higgs-doublet model.
\bigskip
\begin{center}
Table 1
\medskip

\begin{tabular}{lc}
$ht\bar t$ & \Large $\frac{\cos\alpha}{\sin\beta}$ \\ \\
$Ht\bar t$ & \Large $\frac{\sin\alpha}{\sin\beta}$ \\ \\
$At\bar t$ & \Large $\frac{1}{\tan\beta}$ \\ \\
$H^+ \bar t b$ & \Large $\frac{1}{\tan\beta}$
\end{tabular}
\end{center}
\bigskip

\leftline{\large\bf Figure Captions}
\medskip

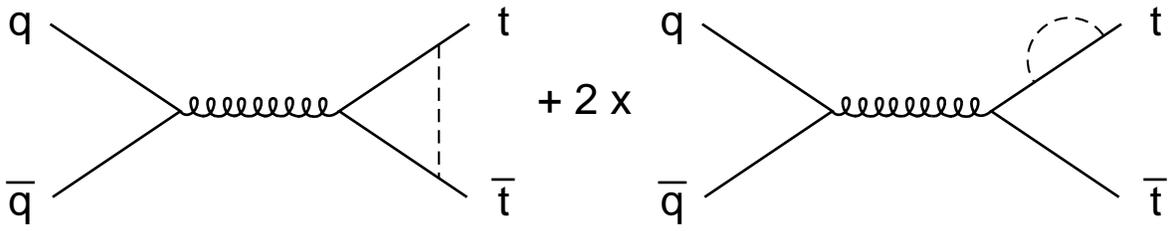
\item[Fig.~1.] Diagrams which contribute to the Yukawa correction to
$\qbq\to\tbt$. The dashed lines represent the Higgs boson and the
unphysical scalar $Z$ and $W$ bosons in Landau gauge.
\bigskip

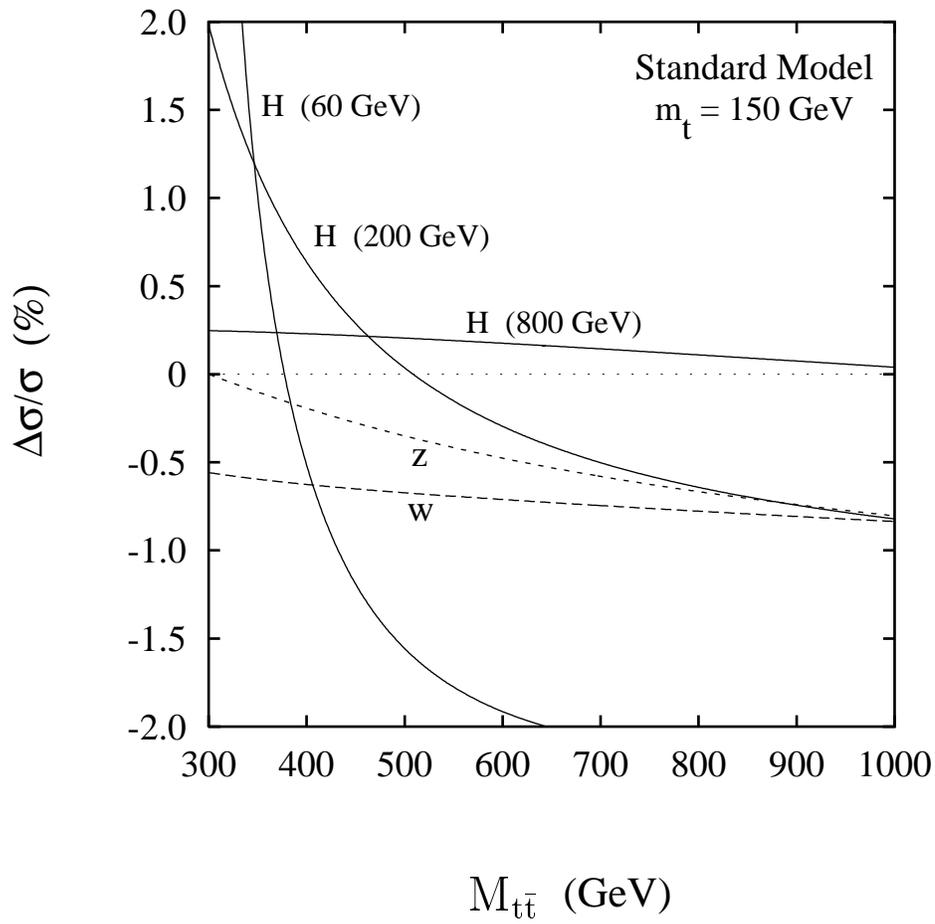
\item[Fig.~2.] Change in the cross section for $q\bar q\to \tbt$, due
to the Yukawa correction, as a function of the \tbt\ invariant mass,
for $m_t=150$ GeV.
The contributions of the Higgs boson and the unphysical scalar
$Z$ and $W$ bosons are shown separately.
\bigskip

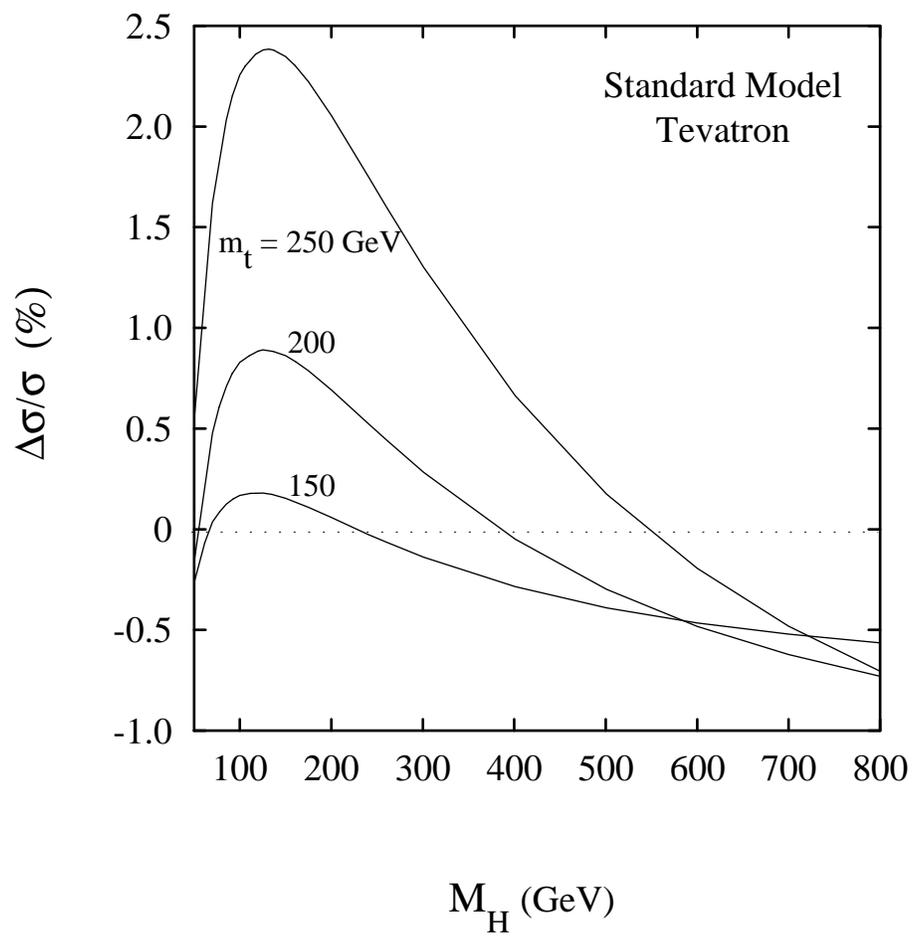
\item[Fig.~3.] Change in the total cross section, due to the Yukawa
correction, for $p\bar p\to \tbt +X$ at the Tevatron, versus the Higgs-boson
mass.
\bigskip

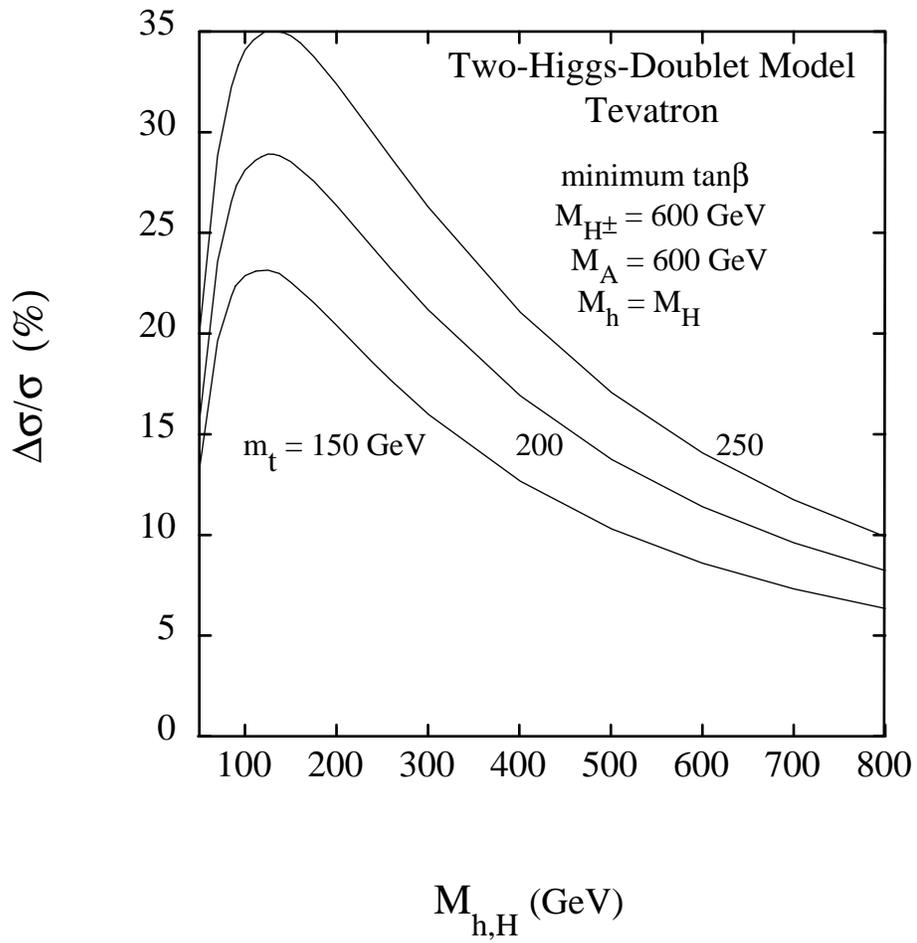
\item[Fig.~4.] Maximum change in the total cross section, due to the Yukawa
correction in a two-Higgs-doublet model,
for $p\bar p\to \tbt +X$ at the Tevatron, versus the common scalar-Higgs-boson
mass.
\bigskip

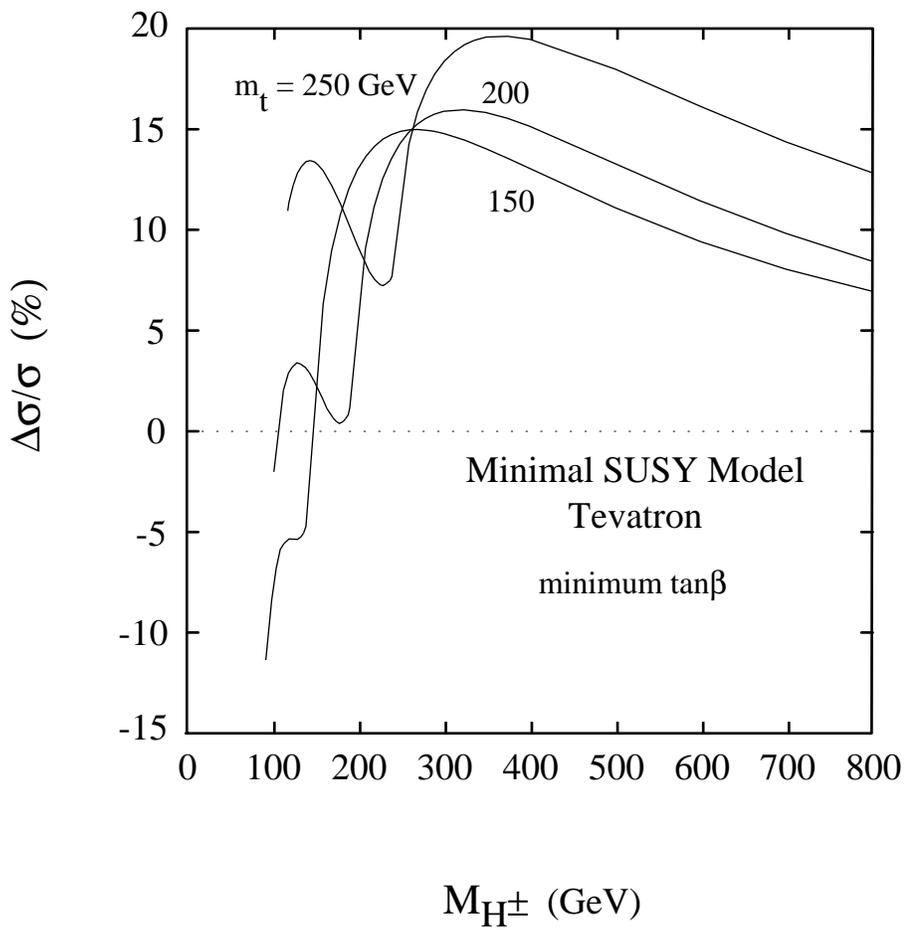
\item[Fig.~5.] Same as Fig.~4, but in the minimal
supersymmetric two-Higgs-doublet
model, and versus the charged-Higgs-boson mass.
\bigskip
\end{description}

\end{document}